\documentclass{article}
\usepackage{waspaa17,amsmath,graphicx,url,times}
\usepackage{amssymb}
\usepackage{color,multirow}

\definecolor{edited_v1}{RGB}{255,0,0}

\title{PSD estimation of multiple sound sources in a reverberant room using a spherical microphone array}

\name{Abdullah Fahim, Prasanga N. Samarasinghe, Thushara D. Abhayapala \thanks{This work is supported by Australian Research Council (ARC) Discovery Projects funding scheme (project no. DP140103412).}
}
\address{Research School of Engineering, The Australian National University
}
\begin{document}
\ninept
\maketitle
\begin{sloppy}
\begin{abstract}
We propose an efficient method to estimate source power spectral densities (PSDs) in a multi-source reverberant environment using a spherical microphone array. The proposed method utilizes the spatial correlation between the spherical harmonics (SH) coefficients of a sound field to estimate source PSDs. The use of the spatial cross-correlation of the SH coefficients allows us to employ the method in an environment with a higher number of sources compared to conventional methods. Furthermore, the orthogonality property of the SH basis functions saves the effort of designing specific beam patterns of a conventional beamformer-based method. We evaluate the performance of the algorithm with different number of sources in practical reverberant and non-reverberant rooms. We also demonstrate an application of the method by separating source signals using a conventional beamformer and a Wiener post-filter designed from the estimated PSDs.
\end{abstract}
\section{Introduction}
\label{sec:intro}
The power spectral density (PSD) of an audio signal carries useful information about the signal characteristics. The information of the source PSD is a desirable quantity required in many speech enhancement techniques, most commonly in Wiener filtering \cite{benesty2005study}. In this work, we use a spherical microphone array (SMA) to estimate the individual source PSDs in a reverberant environment with multiple concurrent speakers and use that information in restoring the original source signals.\par
Hioka et al. proposed a multi-source PSD estimation technique with multiple beamformers (BFs) combining the directivity gains \cite{hioka2013underdetermined}. The authors designed the BFs in an empirical manner which is susceptible to estimation errors due to the ill-conditioning of the demixing matrix. Niwa et al. used the property of an M-matrix to design the BFs in order to improve the condition of the demixing matrix and hence the estimation accuracy \cite{niwa2016psd}. However, in both of the aforementioned cases, the authors considered a non-reverberant environment. Hioka et al. used the spatial correlation between the received signals to propose an alternate method of estimating direct and reverberant signal PSDs for a single source \cite{hioka2011estimating}. The authors of \cite{thiergart2014power} proposed a single source PSD estimator in a diffused sound field using multiple directional microphones.\par
Beamforming is a common speech enhancement technique used for decades \cite{johnson1992array,Bourgeois2010beamforming}. The knowledge of the source PSDs can be used to design a Wiener post-filter at the beamformer (BF) output to enhance the system performance by boosting the interference rejection \cite{marro1998analysis}. Such a combination is used in \cite{hioka2013underdetermined,niwa2016psd} to achieve the source separation in a non-reverberant environment. However, none of the methods discussed above opted for a modal domain solution.\par
The advantage of signal representation in the modal domain \cite{williams1999fourier,abhayapala1999modal} is the inherent orthogonality of their basis functions in terms of spherical harmonics (SH). This orthogonal property allows us to formulate the demixing matrix without the requirement of an explicit design of the BFs. The harmonics sound field coefficients can be recorded using an SMA, or other array structures capable of extracting SH coefficients \cite{abhayapala2002theory,chen2015theory}. Hence, the SH decomposition is becoming a popular tool in the acoustics signal processing such as source localization \cite{evers2014multiple}, speech dereverberation \cite{yamamoto2016spherical}, noise suppression \cite{jarrett2014noise} and beamforming \cite{shabtai2014generalized}. Samarasinghe et al. used the spatial cross-correlation between the sound field coefficients to estimate the PSDs of the direct and reverberant components of a speech signal \cite{samarasinghe2017estimating}. Kalkur et al. proposed a join source localization and separation method in the SH domain using a splitting method based on the Bregman iteration for a non-reverberant case \cite{kalkur2015joint}.\par
The main contribution of this paper is the estimation of source PSDs in a multi-source reverberant environment using SH decomposition. The formulation in the SH domain saves us the extra effort of designing specific BFs \cite{niwa2016psd} by virtue of the inherent orthogonality of the SH. Furthermore, the use of the cross-correlation between the coefficients allows us to separate a higher number of sources compared to the conventional beamforming-based techniques \cite{hioka2013underdetermined,niwa2016psd}. The estimated PSD is used in a two-step source separation algorithm to demonstrate an application of the method. We use a commercially available higher order microphone (HOM) \lq{Eigenmike}\rq \cite{eigenmikeweb} to evaluate the performance of the algorithm in different practical environments.
\section{Problem formulation}
\label{sec:problem_formulation}
Let us consider an SMA to capture the sound field generated by $L$ uncorrelated far-field sources in a reverberant room. We assume that the SMA consists of $Q$ pressure microphones and denote the position of the $q^{th}$ microphone by $\boldsymbol{x}_q = (r, \theta_q, \phi_q)$, where $q \in [1, Q]$. The received signal at the $q^{th}$ microphone is
\begin{multline} \label{eq:basic-reverb-model}
P(\boldsymbol{x_q}, k) = \sum \limits_{\ell=1}^{L} S_{\ell}(k) \bigg[ G^{(\ell)}_{d}(k) e^{ik \text{ } \boldsymbol{\hat{y}_{\ell}} \cdot \boldsymbol{x_q}} + \\
\int_{\boldsymbol{\hat{y}}} G^{(\ell)}_r(k, \boldsymbol{\hat{y}}) e^{ik \text{ } \boldsymbol{\hat{y}} \cdot \boldsymbol{x_q}} \text{ } d\boldsymbol{\hat{y}} \bigg]
\end{multline}
where $k = 2 \pi f / c$, $f$ is the frequency, $c$ is the speed of sound propagation, $\boldsymbol{\hat{y}_{\ell}}$ is a unit vector towards the direction of the $\ell^{th}$ source, $G^{(\ell)}_r(k, \boldsymbol{\hat{y}})$ is the reflection gain along an arbitrary the direction of $\boldsymbol{\hat{y}}$ for the $\ell^{th}$ source, and $S_{\ell}(k)$ and $G^{(\ell)}_{d}(k)$ represent the source strength and the direct path gain for the $\ell^{th}$ source, respectively. Given the measured sound pressure $P(\boldsymbol{x_q}, k)$, we aim to estimate the PSD of each source signal $S_{\ell}(k)$ and separate the individual sources.
\section{PSD estimation}
\label{sec:psd_estimation}
The SH decomposition of an $N^{th}$ order sound field is given by \cite[ch. 6]{williams1999fourier}
\begin{equation} \label{eq:basic-spherical-harmonics-equation}
P(\boldsymbol{x_q}, k) = \sum \limits_{n=0}^{N} \sum \limits_{m=-n}^{n} \alpha_{nm}(k) \text{ } b_n(kr) \text{ } Y_{nm}(\theta_q, \phi_q)
\end{equation}
where $N = \lceil kr \rceil$ \cite{ward2001reproduction}, $\lceil \cdot \rceil$ denotes the ceiling operation, $Y_{nm}(\cdot)$ is the SH function of order $n$ and degree $m$, and 
\begin{equation} \label{eq:bn}
b_n(kr) = 
\begin{cases}
j_n(kr) & \text{for an open array} \\
j_n(kr) - \frac{j'_n(kr)}{h'_n(kr)} h_n(kr) & \text{for a rigid array}
\end{cases}
\end{equation}
with $j_n(\cdot)$ and $h_n(\cdot)$ denoting the $n^{th}$ order spherical Bessel and Hankel functions, respectively, and $(\cdot)'$ refers to the first derivative. Utilizing the orthogonal property of the SH, the sound field coefficients $\alpha_{nm}(k)$ can be calculated using an SMA by \cite{samarasinghe20123d}
\begin{equation} \label{eq:alpha}
\alpha_{nm}(k) = \frac{1}{b_n(k)} \sum \limits_{q=1}^{Q} P(\boldsymbol{x_q}, k) \text{ } Y_{nm}^*(\theta_q, \phi_q)
\end{equation}
where $*$ denotes the complex conjugate operation. Furthermore, a SH based solution for the sound field due to a far-field unit amplitude source is given by \cite[pp. 9--13]{teutsch2007modal}
\begin{equation} \label{eq:far-field-source}
e^{ik \text{ } \boldsymbol{\hat{y}_{\ell}} \cdot \boldsymbol{x_q}} = \sum \limits_{n=0}^{N} \sum \limits_{m=-n}^{n} 4 \pi i^{n} \text{ } Y^*_{nm}(\boldsymbol{\hat{y}_{\ell}}) \text{ } b_n(kr) \text{ } Y_{nm}(\theta_q, \phi_q).
\end{equation}
Using \eqref{eq:basic-spherical-harmonics-equation} and \eqref{eq:far-field-source} in \eqref{eq:basic-reverb-model}, we derive
\begin{multline} \label{eq:alpha-basic-model}
\alpha_{nm}(k) = \sum \limits_{\ell=1}^{L} 4 \pi i^{n} \text{ } S_{\ell}(k)  \bigg[ G_d^{(\ell)}(k) \text{ } Y^*_{nm}(\boldsymbol{\hat{y}_{\ell}}) + \\ \int_{\boldsymbol{\hat{y}}} G^{(\ell)}_r(k, \boldsymbol{\hat{y}}) \text{ } Y^*_{nm}(\boldsymbol{\hat{y}}) \text{ } d\boldsymbol{\hat{y}} \bigg].
\end{multline}
From \eqref{eq:alpha-basic-model}, the spatial correlation between $\alpha_{nm}(k)$ and $\alpha_{n'm'}(k)$ is
\begin{multline} \label{eq:cross-alpha-1}
E\left\{ \alpha_{nm}(k) \alpha^*_{n'm'}(k) \right\} = C_{nn'} \sum \limits_{\ell=1}^{L} \sum \limits_{\ell'=1}^{L}  \text{ } E\{S_{\ell}(k) \text{ } S^*_{\ell'}(k)\} \\
\times E \Bigg\{ \left[ G_d^{(\ell)}(k) \text{ } Y^*_{nm}(\boldsymbol{\hat{y}_{\ell}}) + \int_{\boldsymbol{\hat{y}}} G^{(\ell)}_r(k, \boldsymbol{\hat{y}}) \text{ } Y^*_{nm}(\boldsymbol{\hat{y}}) \text{ } d\boldsymbol{\hat{y}} \right] \times \\
\left[ G_d^{(\ell')*}(k) \text{ } Y_{n'm'}(\boldsymbol{\hat{y}_{\ell'}}) + \int_{\boldsymbol{\hat{y}'}} G_r^{(\ell')*}(k, \boldsymbol{\hat{y}'}) \text{ } Y_{n'm'}(\boldsymbol{\hat{y}'}) \text{ } d\boldsymbol{\hat{y}'} \right] \Bigg\}
\end{multline}
where $C_{nn'} \triangleq 16 \pi^2 i^{n} (-i)^{n'}$ and $E\{\cdot\}$ represents the expected value over time. Due to the autonomous behavior of the reflective surfaces in a room (i.e., the reflection gains from the reflective surfaces are independent from the direct path gain), the cross correlation between the direct path gain and reverberant path gain coefficients can be assumed to be negligible, e.g.,
\begin{equation} \label{eq:uncorrelated-reverb-direct}
E\{ G_d^{(\ell)}(k) \text{ } G_r^{(\ell)*}(k, \boldsymbol{\hat{y}}) \} = 0.
\end{equation}
We assume that the sources are uncorrelated with each other, and so do the reverberant path gains from different directions, e.g.
\begin{equation} \label{eq:uncorrelated-direct-sound}
E\{ S_\ell(k) \text{ } S^*_{\ell'}(k) \} = E\{ \lvert S_\ell(k) \rvert ^2 \} \text{ } \delta_{\ell \ell'}
\end{equation}
\begin{equation} \label{eq:uncorrelated-reverb-sound}
E\{ G_r^{(\ell)}(k, \boldsymbol{\hat{y}}) \text{ } G_r^{(\ell)*}(k, \boldsymbol{\hat{y}'}) \} = \lvert G_r^{(\ell)}(k, \boldsymbol{\hat{y}}) \rvert ^2 \text{ } \delta_{\boldsymbol{\hat{y}} \boldsymbol{\hat{y}'}}
\end{equation}
where $\delta_{\boldsymbol{\hat{y}} \boldsymbol{\hat{y}'}}$ and $\delta_{\ell \ell'}$ are the Kronecker delta functions and $\lvert \cdot \rvert$ denotes the absolute value. Using \eqref{eq:uncorrelated-reverb-direct}, \eqref{eq:uncorrelated-direct-sound} and \eqref{eq:uncorrelated-reverb-sound} in \eqref{eq:cross-alpha-1}, we get
\begin{multline}\label{eq:cross-alpha-2}
E\{ \alpha_{nm}(k) \alpha^*_{n'm'}(k) \} = C_{nn'} \sum \limits_{\ell=1}^{L} \Bigg[ \Phi_\ell(k) \text{ } Y^*_{nm}(\boldsymbol{\hat{y}_{\ell}}) \text{ } Y_{n'm'}(\boldsymbol{\hat{y}_{\ell}}) \\
+\int_{\boldsymbol{\hat{y}}} E\{\lvert S_{\ell}(k) \rvert ^2\} E\{ \lvert G^{(\ell)}_r(k, \boldsymbol{\hat{y}}) \rvert ^2 \} Y^*_{nm}(\boldsymbol{\hat{y}}) Y_{n'm'}(\boldsymbol{\hat{y}}) d\boldsymbol{\hat{y}} \Bigg]
\end{multline}
where $\Phi_\ell(k)=E\{\lvert S_{\ell}(k) \rvert ^2\} \text{ } E\{ \lvert G_d^{(\ell)}(k) \rvert ^2 \}$ is the PSD of the $\ell^{th}$ source. Since $\lvert G^{(\ell)}_r(k, \boldsymbol{\hat{y}}) \rvert ^2$ is defined over a sphere, we can represent it in terms of a SH decomposition as
\begin{equation} \label{eq:spherical-power}
E\{ \lvert G^{(\ell)}_r(k, \boldsymbol{\hat{y}}) \rvert ^2 \} = \sum \limits_{v=0}^{V} \sum \limits_{u=-v}^{v} \gamma^{(\ell)}_{vu}(k) \text{ } Y_{vu}(\boldsymbol{\hat{y}})
\end{equation}
where $V$ is the harmonics order, which theoretically extends to the infinity. However, in practice, we limit $V$ to an empirically decided value to keep the system well-conditioned. Substituting the value of $E\{ \lvert G^{(\ell)}_r(k, \boldsymbol{\hat{y}}) \rvert ^2 \}$ from \eqref{eq:spherical-power} into \eqref{eq:cross-alpha-2}, we derive
\begin{multline}\label{eq:final-model-rev}
\underbrace{ E\{ \alpha_{nm}(k) \alpha^*_{n'm'}(k) \} }_{\triangleq\Lambda_{nm}^{n'm'}(k)} = \sum \limits_{\ell=1}^{L} \Phi_\ell(k) \text{ } \underbrace{ C_{nn'} Y^*_{nm}(\boldsymbol{\hat{y}_{\ell}}) \text{ } Y_{n'm'}(\boldsymbol{\hat{y}_{\ell}}) }_{\triangleq \Upsilon_{nm}^{n'm'}(\boldsymbol{\hat{y}_\ell})} \\
+ \sum \limits_{v=0}^{V} \sum \limits_{u=-v}^{v} \Gamma_{vu}(k) \underbrace{C_{nn'} \int_{\boldsymbol{\hat{y}}} Y_{vu}(\boldsymbol{\hat{y}})  Y^*_{nm}(\boldsymbol{\hat{y}}) Y_{n'm'}(\boldsymbol{\hat{y}}) d\boldsymbol{\hat{y}} }_{\triangleq \Psi_{n,n',v}^{m,m',u} = C_{nn'} \text{ }W_{n,n',v}^{m,m',u}}
\end{multline}
where $\Gamma_{vu}(k) \triangleq \sum \limits_{\ell=1}^{L} \gamma^{(\ell)}_{vu}(k) \text{ } E\{\lvert S_{\ell}(k) \rvert ^2\}$ and from the integral property of the SH
\begin{equation} \label{eq:wigner}
W_{n,n',v}^{m,m',u} = (-1)^m \text{ } \sqrt[]{\frac{(2v+1)(2n+1)(2n'+1)}{4 \pi}} \text{ }W_{12}
\end{equation}
with $W_{12}$ representing a multiplication between two Wigner-3j symbols \cite{olver2010nist} as
\begin{equation} \label{eq:w12}
W_{12} = \left(\begin{array}{clcr}
v & n & n'\\
0 & 0 & 0  \end{array}\right) \text{ }
\left(\begin{array}{clcr}
v & n & n'\\
u & -m & m'  \end{array}\right).
\end{equation}
Considering the cross-correlation of all the available SH coefficients, \eqref{eq:final-model-rev} can be written in a matrix form as
\begin{equation} \label{eq:final-model-rev-matrix}
\boldsymbol{\Lambda} = \boldsymbol{T} \text{ } \boldsymbol{\Theta}
\end{equation}
where
\begin{equation} \label{eq:Lambda}
\boldsymbol{\Lambda} = [\Lambda_{00}^{00} \quad \Lambda_{00}^{1-1} \dots \Lambda_{00}^{NN} \quad \Lambda_{1-1}^{00} \dots \Lambda_{NN}^{NN}]^T_{1 \times (N+1)^4}
\end{equation}
\begin{equation} \label{eq:T}
\boldsymbol{T} = 
\underbrace{
\begingroup % keep the change local
\setlength\arraycolsep{2pt}
\begin{bmatrix}
    \Upsilon_{00}^{00}(\boldsymbol{\hat{y}_1}) & \dots  &  {\Upsilon_{00}^{00}(\boldsymbol{\hat{y}_L})}  & \Psi_{0,0,0}^{0,0,0} & \dots & \Psi_{0,0,V}^{0,0,V} \\
    \vdots & \vdots & \vdots & \vdots & \vdots & \vdots \\
    \vdots & \vdots & \vdots & \vdots & \vdots & \vdots \\
    \Upsilon_{NN}^{NN}(\boldsymbol{\hat{y}_1}) & \dots  &  {\Upsilon_{NN}^{NN}(\boldsymbol{\hat{y}_L})}  & \Psi_{N,N,0}^{N,N,0} & \dots & \Psi_{N,N,V}^{N,N,V} \\
  \end{bmatrix} 
  \endgroup
}_{(N+1)^4 \times (L+\{V+1\}^2)}
\end{equation}
\begin{equation} \label{eq:Psi}
\boldsymbol{\Theta} = [\Phi_1 \dots \Phi_{L} \quad \Gamma_{00} \dots \Gamma_{VV}]^T_{1 \times (L+\{V+1\}^2)}.
\end{equation}
Note that, the frequency dependency is omitted in \eqref{eq:Lambda}-\eqref{eq:Psi} to simplify the notation. For practical implementation, the expected value $\Lambda_{nm}^{n'm'}(k)$ is estimated using an exponentially weighted moving average as
\begin{equation} \label{eq:expected-value}
\Lambda_{nm}^{n'm'}(\tau, k) = \beta \text{ } \Lambda_{nm}^{n'm'}(\tau-1, k) + (1 - \beta) \text{ } \alpha_{nm}(\tau, k) \alpha^*_{n'm'}(\tau, k)
\end{equation}
where $\beta \in [0, 1]$ is a smoothing factor, $\tau$ denotes the time frame index in the short time Fourier transform (STFT) domain, and $k$ is calculated from the center frequency of the corresponding STFT bin. Hence, the source and reverberant PSDs are estimated by
\begin{equation} \label{eq:solution}
\boldsymbol{\hat{\Theta}} = \boldsymbol{T}^{\dagger} \text{ } \boldsymbol{\Lambda}
\end{equation}
where $^{\dagger}$ indicates the pseudo-inverse operation. In the practical implementation, a half-wave rectification is performed on \eqref{eq:solution} to avoid negative PSDs. It is worth noting that, \eqref{eq:solution} can readily be used for estimating source PSDs in a non-reverberant environment by discarding the $\Psi$ terms from the translation matrix $\boldsymbol{T}$ in \eqref{eq:T}.
\section{Application in source separation}
We use a BF and a Wiener post-filter to separate the source signals \cite{niwa2016psd} in a multi-source reverberant environment. The choice of the BF can vary based on the specific design criteria. In our work, we use a maximum directivity BF formulated in the SH domain \cite{meyer2002highly}.
\subsection{Maximum directivity beamformer}
The output of a maximum directivity BF steered towards $\ell^{th}$ far-field source is given by \cite{meyer2002highly,rafaely2008spherical}
\begin{equation} \label{eq:bf-output}
Z_{\ell}(k) = \sum \limits_{n=0}^{N} \sum \limits_{m=-n}^{n} \frac{i^{-n}}{(N+1)^2} \alpha_{nm}(k) Y_{nm}(\theta_{\ell}, \phi_{\ell}).
\end{equation}
Equation \eqref{eq:bf-output} requires the knowledge of the source directions which can be estimated using any suitable localization algorithm.
\subsection{Wiener post-filter}
The total reverberant power due to all the sources is
\begin{equation} \label{eq:reverb-power}
\Phi_r(k) = \sum \limits_{\ell=1}^{L} E\{\lvert S_{\ell}(k) \rvert ^2\} \int_{\boldsymbol{\hat{y}}} E\{ \lvert G^{(\ell)}_r(k, \boldsymbol{\hat{y}}) \rvert ^2 \} \text{ } d\boldsymbol{\hat{y}}.
\end{equation}
Using \eqref{eq:spherical-power}, the definition of $\Gamma_{vu}(k)$ and the symmetrical property of the SH, \eqref{eq:reverb-power} can be written as
\begin{align} \label{eq:reverb-power-2}
\Phi_r(k) & = \sum \limits_{v=0}^{V} \sum \limits_{u=-v}^{v} \Gamma_{vu}(k) \text{ } \int_{\boldsymbol{\hat{y}}} Y_{vu}(\boldsymbol{\hat{y}}) \text{ } d\boldsymbol{\hat{y}} \nonumber \\
& = \sum \limits_{v=0}^{V} \sum \limits_{u=-v}^{v} \Gamma_{vu}(k) \text{ } \frac{\delta(v) \delta(u)}{\sqrt[]{4\pi}} \nonumber \\
& = \frac{\Gamma_{00}(k)}{\sqrt[]{4\pi}}
\end{align}
where $\delta(\cdot)$ is the Dirac delta function. Hence, applying the Wiener filter at the BF output, we estimate the $\ell^{th}$ source strength by
\begin{equation} \label{eq:final-estimation}
\hat{S}_{\ell}(k) = Z_{\ell}(k) \text{ } \frac{\Phi_{\ell}(k)}{\sum \limits_{\ell'=1}^{L} \Phi_{\ell'}(k) + \Phi_r(k)}.
\end{equation}
\begin{figure}[t]
\begin{minipage}[b]{0.5\linewidth}
  \centering
  \centerline{\includegraphics[width=\linewidth]{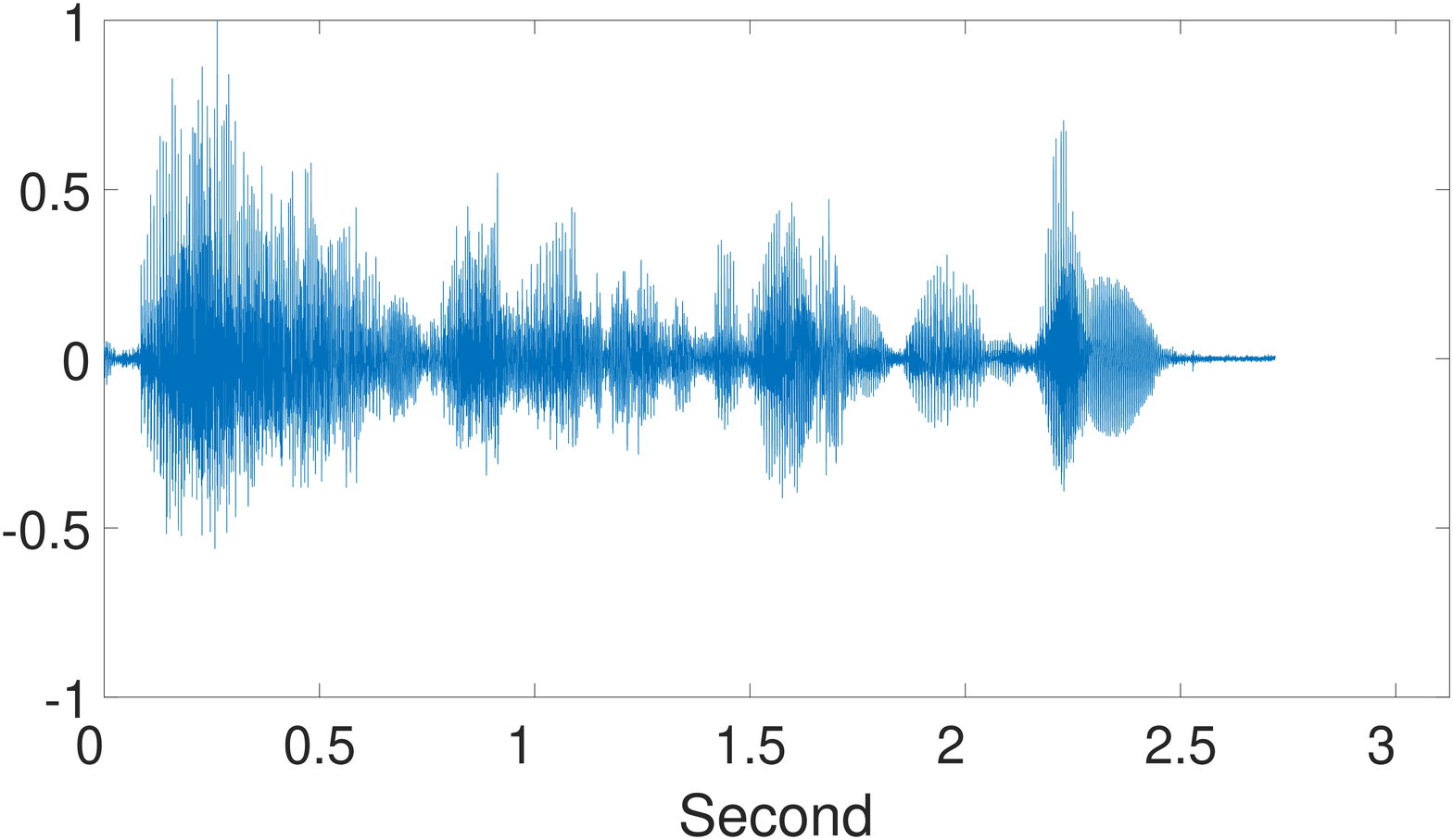}}
  \centerline{(a) Mixed signal}\medskip
\end{minipage}
\hfill
\begin{minipage}[b]{.5\linewidth}
  \centering
  \centerline{\includegraphics[width=\linewidth]{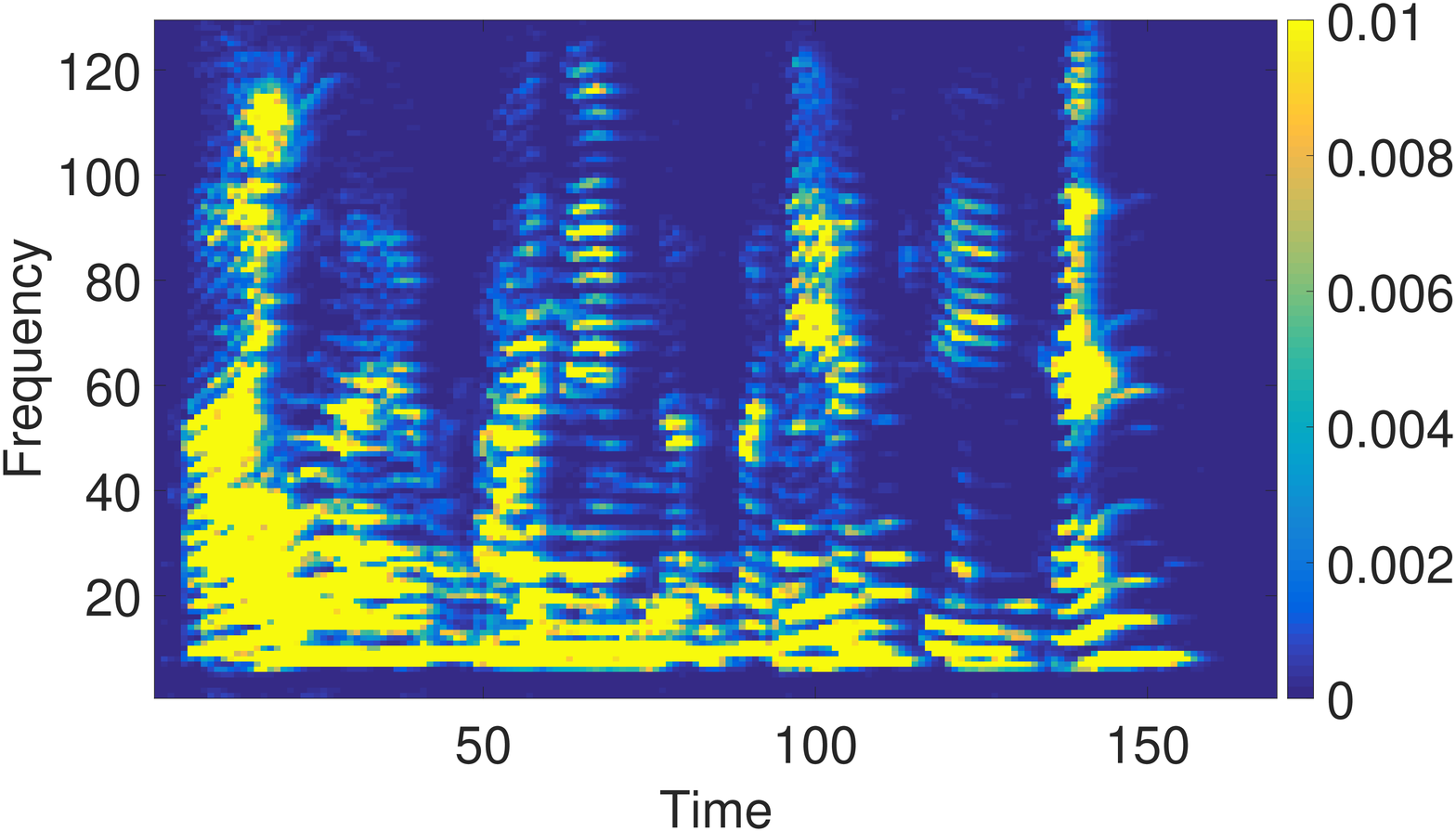}}
  \centerline{(d) Mixed PSD}\medskip
\end{minipage}
\begin{minipage}[b]{0.5\linewidth}
  \centering
  \centerline{\includegraphics[width=\linewidth]{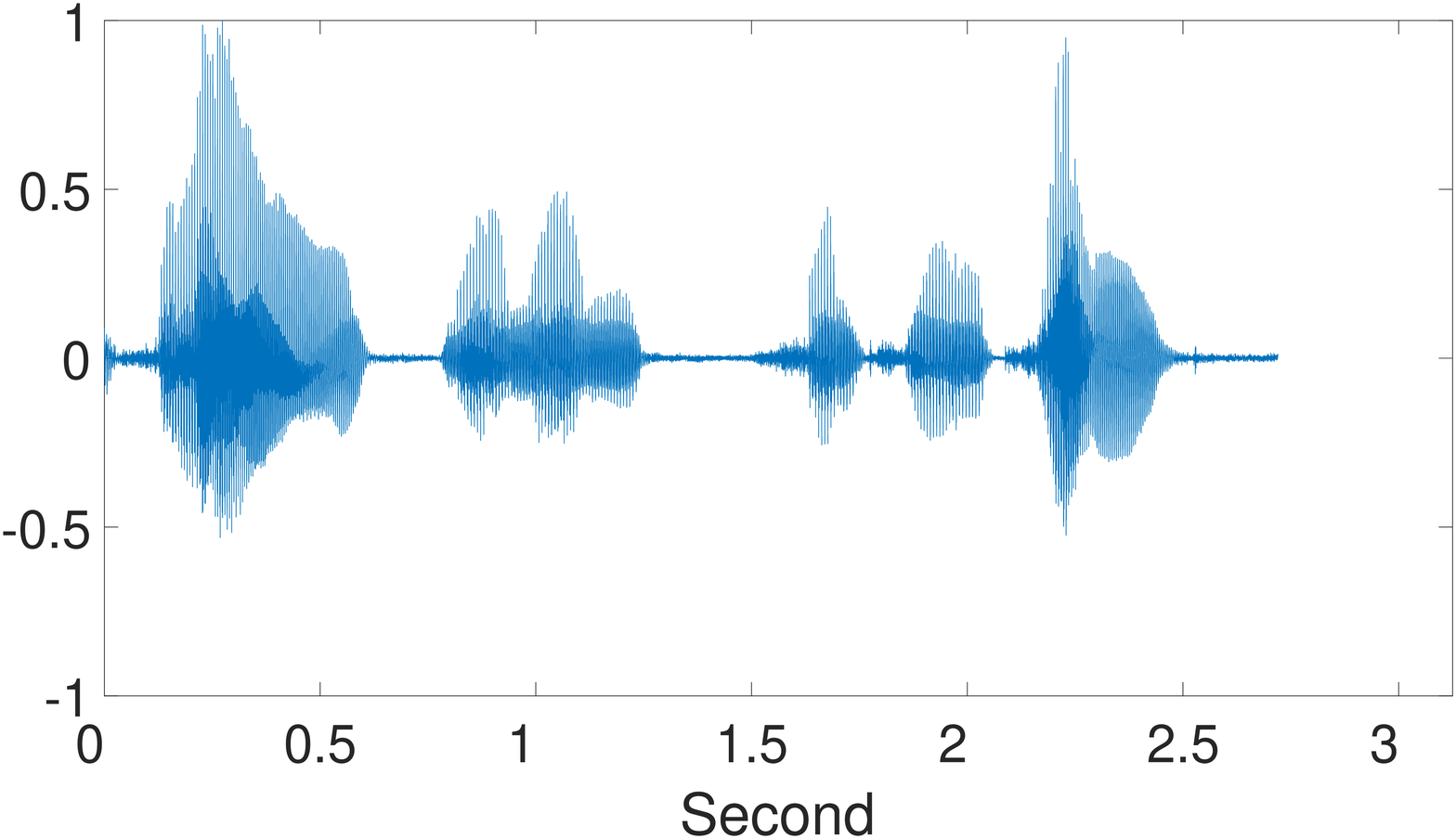}}
  \centerline{(b) Speaker 1 (original)}\medskip
\end{minipage}
\hfill
\begin{minipage}[b]{.5\linewidth}
  \centering
  \centerline{\includegraphics[width=\linewidth]{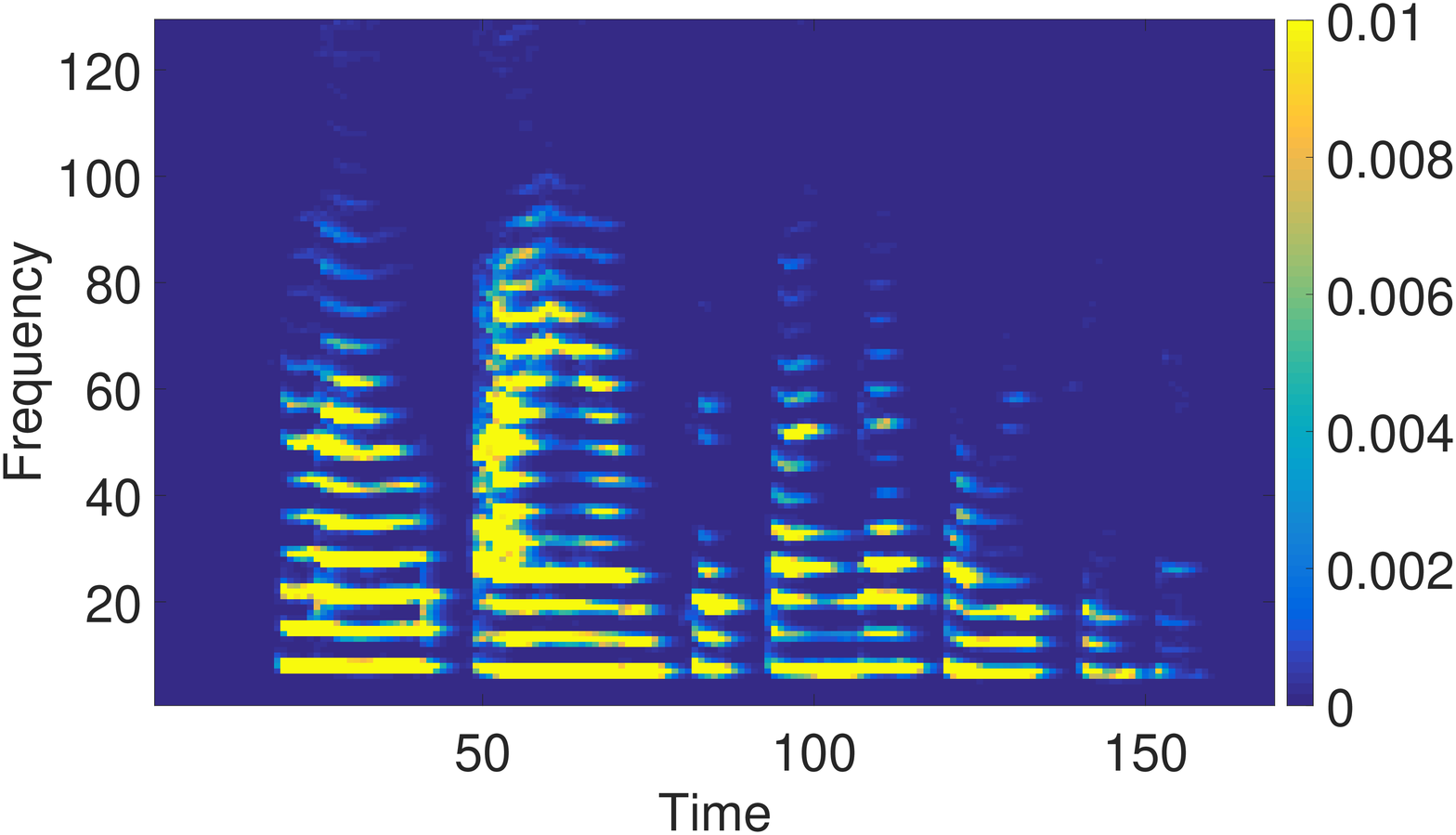}}
  \centerline{(e) Speaker 2 (original)}\medskip
\end{minipage}
\begin{minipage}[b]{0.5\linewidth}
  \centering
  \centerline{\includegraphics[width=\linewidth]{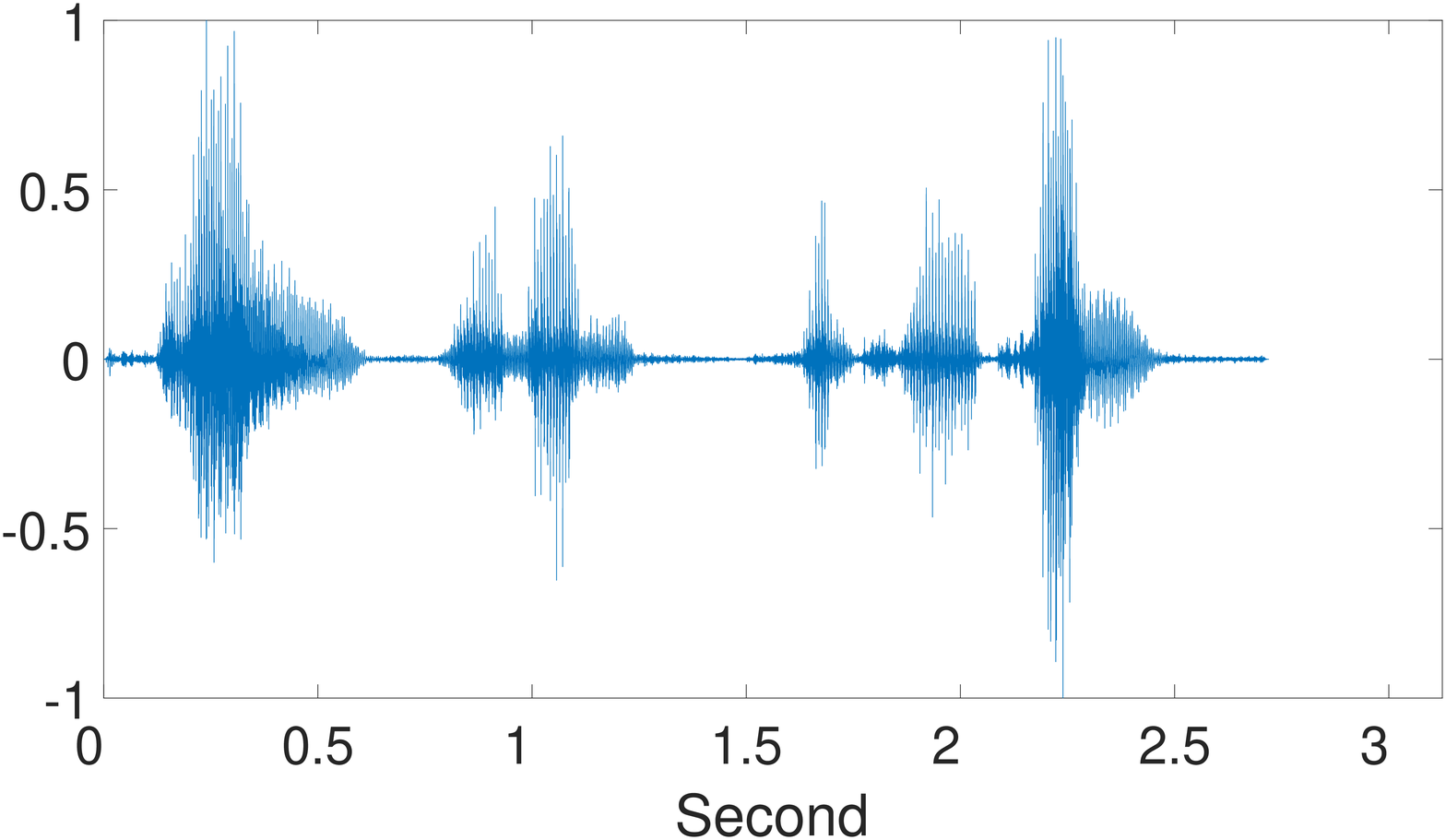}}
  \centerline{(c) Speaker 1 (estimated)}\medskip
\end{minipage}
\hfill
\begin{minipage}[b]{.5\linewidth}
  \centering
  \centerline{\includegraphics[width=\linewidth]{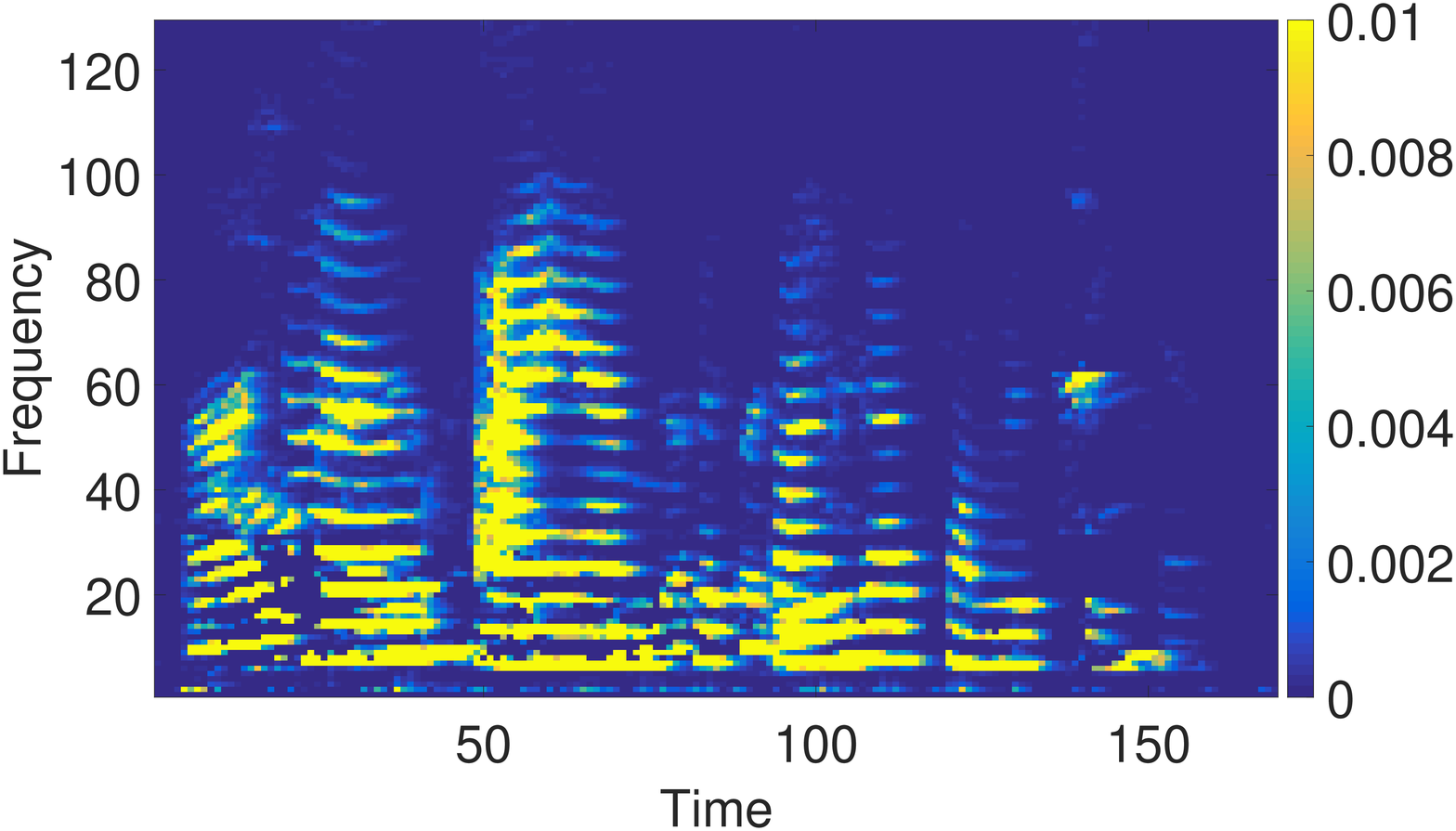}}
  \centerline{(f) Speaker 2 (estimated)}\medskip
\end{minipage}
\caption{Evaluation in an anechoic chamber. Figure (a)-(c) and (d)-(f) show the estimated waveforms and PSDs, respectively.}
\label{fig:anechoic_room}
\end{figure}
\section{Experimental result}
\label{sec:experimental_result}
\subsection{Experimental Setup}
The evaluation of the proposed method in the practical and simulated environments is demonstrated in this section. Data processing was performed in the frequency domain after converting the recorded signals using a $256$-point STFT with a $32$ ms Hanning window and $50\%$ overlap. To reduce the computational cost, all the signals were decimated to $8000$ Hz sampling frequency. We placed the sources at a distance of $2$ m from the center of an SMA which had a radius of $4.2$ cm. Due to a larger source to microphone distance compared to the SMA radius, the sources were considered to be far-field sources. A flat value of $\beta = 0.4$ was used in \eqref{eq:expected-value} for all the experiments.
\subsection{Anechoic chamber}
\label{sec: anechoic-chamber}
We performed the evaluation of the proposed algorithm in an anechoic chamber with $4$ human speakers. The mixed speech signal was recorded with a $4^{th}$ order $32$-channels Eigenmike \cite{eigenmikeweb}. We estimated the directions of arrival (DOA) of the sources (Table \ref{table:doa}) using a frequency-smoothed MUSIC algorithm \cite{khaykin2009coherent}. Fig. \ref{fig:anechoic_room} compares the estimation results for the first two speakers with the respective reference signals. The time domain signal representation of speaker $1$ shows a great resemblance with the original speech signal. The estimated PSD of speaker $2$ also displays a good result except the first few time frames, which is expected due to the fact that we used a moving average across the time frames to estimate the expected values. The informal listening tests also confirm a satisfactory separation performance for all the $4$ speech signals with marginal spectral distortion. The average signal to interference ratio (SIR) \cite{vincent2006performance,vincent2005bss} was calculated as $14.69$ dB.
\subsection{Practical environment}
\label{sec: practical-environment}
We also evaluated the algorithm in a practical reverberant room with a similar setup used in the anechoic chamber. The mixed signal recorded in the Eigenmike was generated by playing $4$ distinct audio signals from the WSJCAM0 corpus \cite{robinson1995wsjcamo} using different speakers. Table \ref{table:doa} shows the estimated DOA of the setup. The separated PSD for source $1$ is plotted in Fig. \ref{fig:lab_room} along with the reference signals. From Fig. \ref{fig:lab_room}(a) and \ref{fig:lab_room}(c), we can observe the similarity between the reference and estimated PSDs. Fig. \ref{fig:lab_room}(d) plots the estimated PSD when  room reflections were ignored, and as expected, we see some distortions and spectral overlapping due to the unaccounted reverberation components. The average SIR for this case was $10.03$ dB.
\begin{table}[ht]
\caption{Source positions $(\theta, \phi)$}
\label{table:doa}
\begin{center}
\renewcommand{\arraystretch}{1.2}
\begin{tabular}{|c|c|c|}
\hline
 & \textbf{Anechoic chamber} & \textbf{Reverberant room} \\
\hline
Source 1 & $(78.01 ^ \circ, 50.42 ^ \circ)$ & $(74.2 ^ \circ, 27.22 ^ \circ)$ \\
\hline
Source 2 & $(77.15 ^ \circ, 141.81 ^ \circ)$ & $(76.78 ^ \circ, 55.58 ^ \circ)$ \\
\hline
Source 3 & $(76.29 ^ \circ, 218.87 ^ \circ)$ & $(77.06 ^ \circ, 87.09 ^ \circ)$ \\
\hline
Source 4 & $(71.42 ^ \circ, 313.69 ^ \circ)$ & $(73.91 ^ \circ, 324.90 ^ \circ)$ \\
\hline
\end{tabular}
\end{center}
\end{table}
\begin{figure}[!ht]
\begin{minipage}[b]{0.5\linewidth}
  \centering
  \centerline{\includegraphics[width=\linewidth]{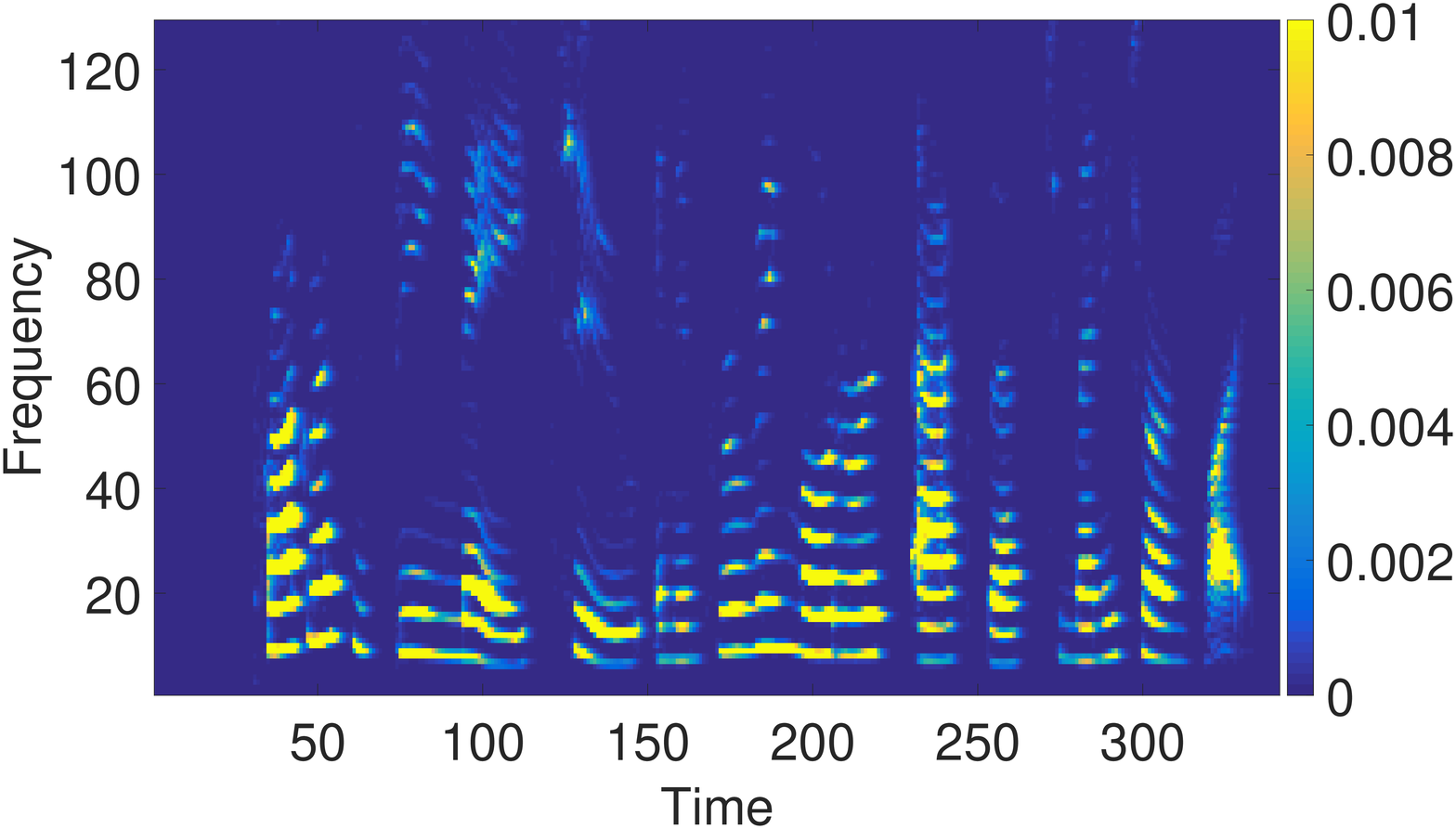}}
  \centerline{(a) Clean PSD}\medskip
\end{minipage}
\hfill
\begin{minipage}[b]{.5\linewidth}
  \centering
  \centerline{\includegraphics[width=\linewidth]{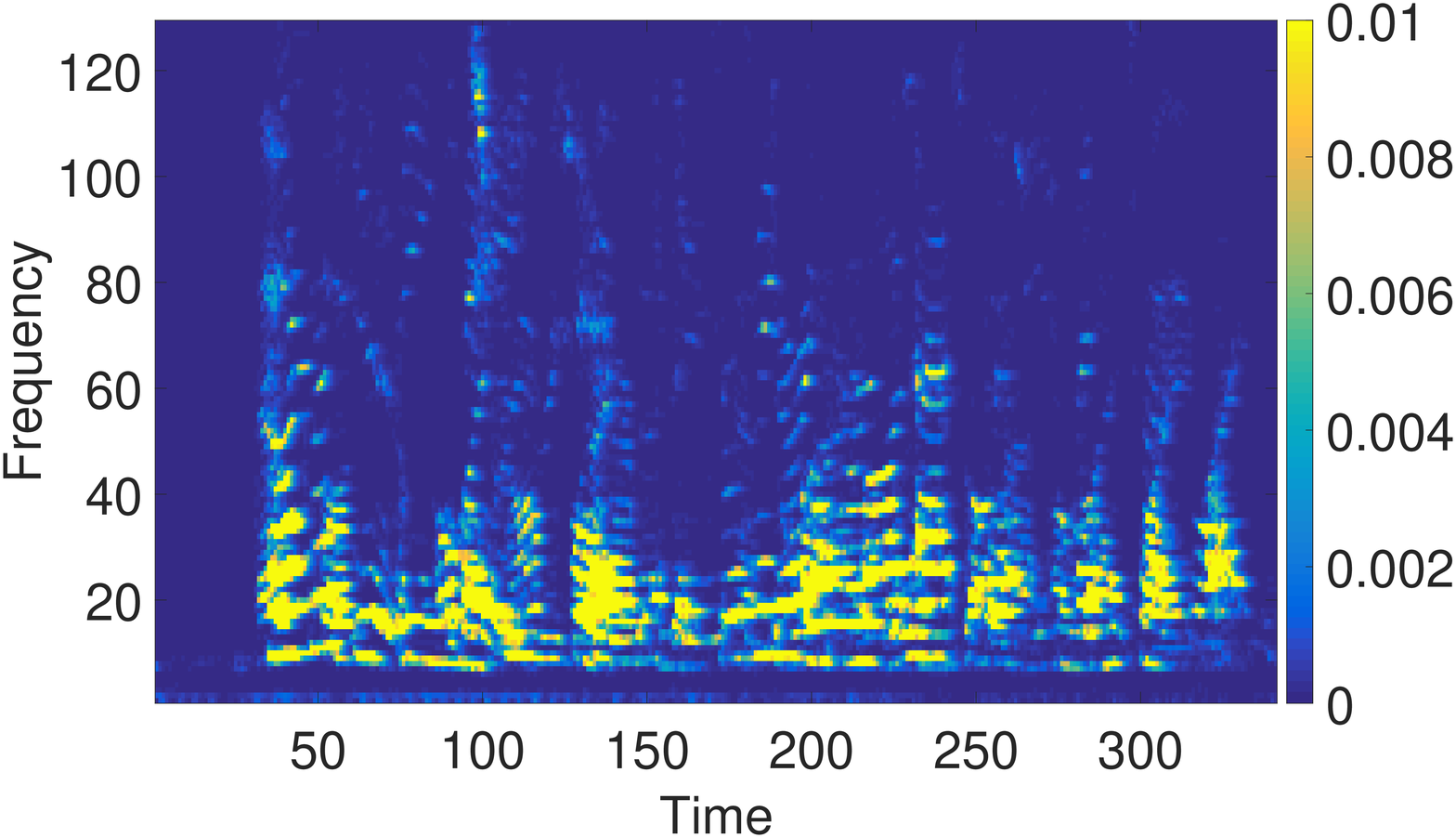}}
  \centerline{(b) Mixed PSD}\medskip
\end{minipage}
\begin{minipage}[b]{0.5\linewidth}
  \centering
  \centerline{\includegraphics[width=\linewidth]{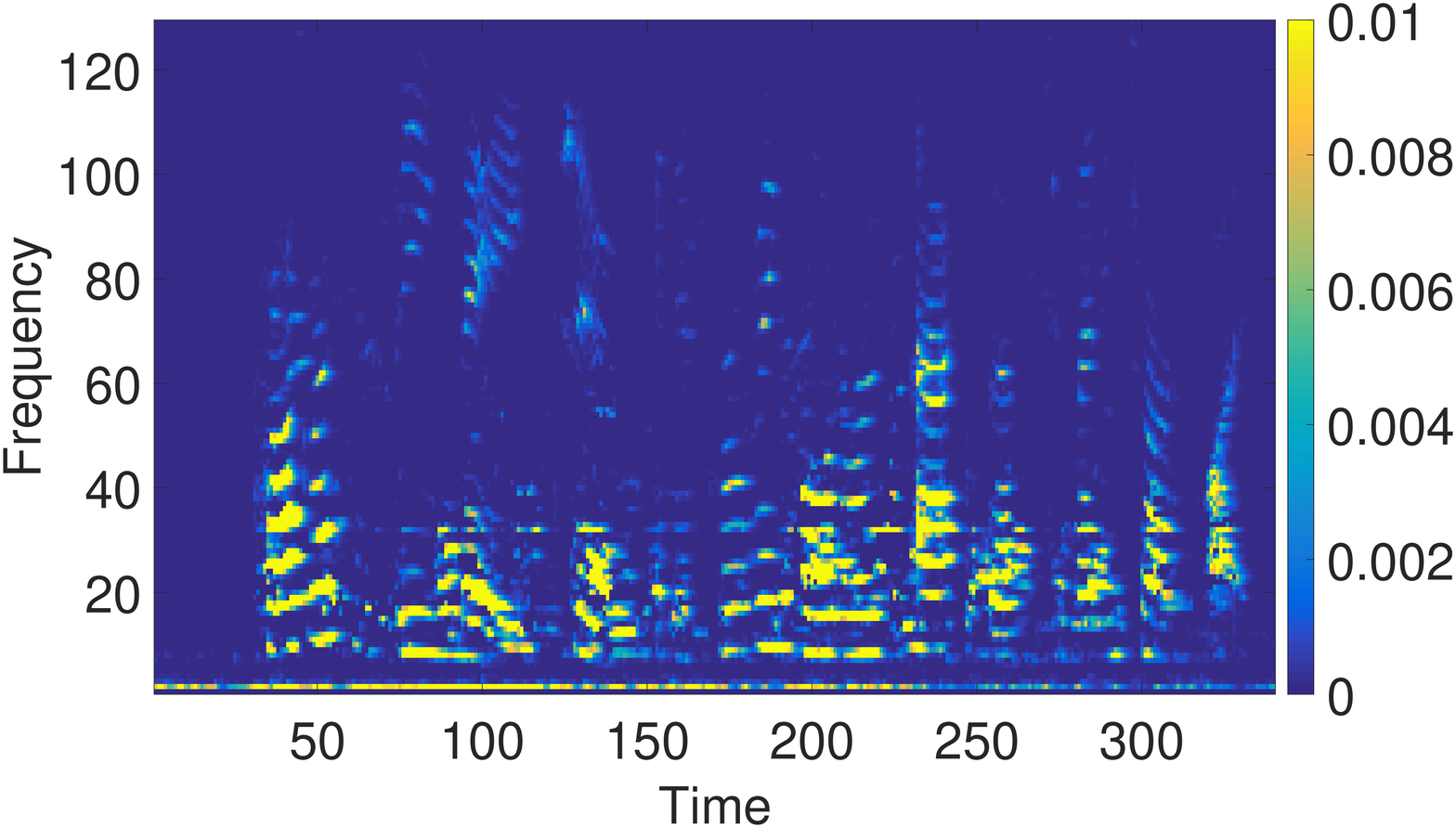}}
  \centerline{(c) Estimated PSD ($V=2$)}\medskip
\end{minipage}
\hfill
\begin{minipage}[b]{.5\linewidth}
  \centering
  \centerline{\includegraphics[width=\linewidth]{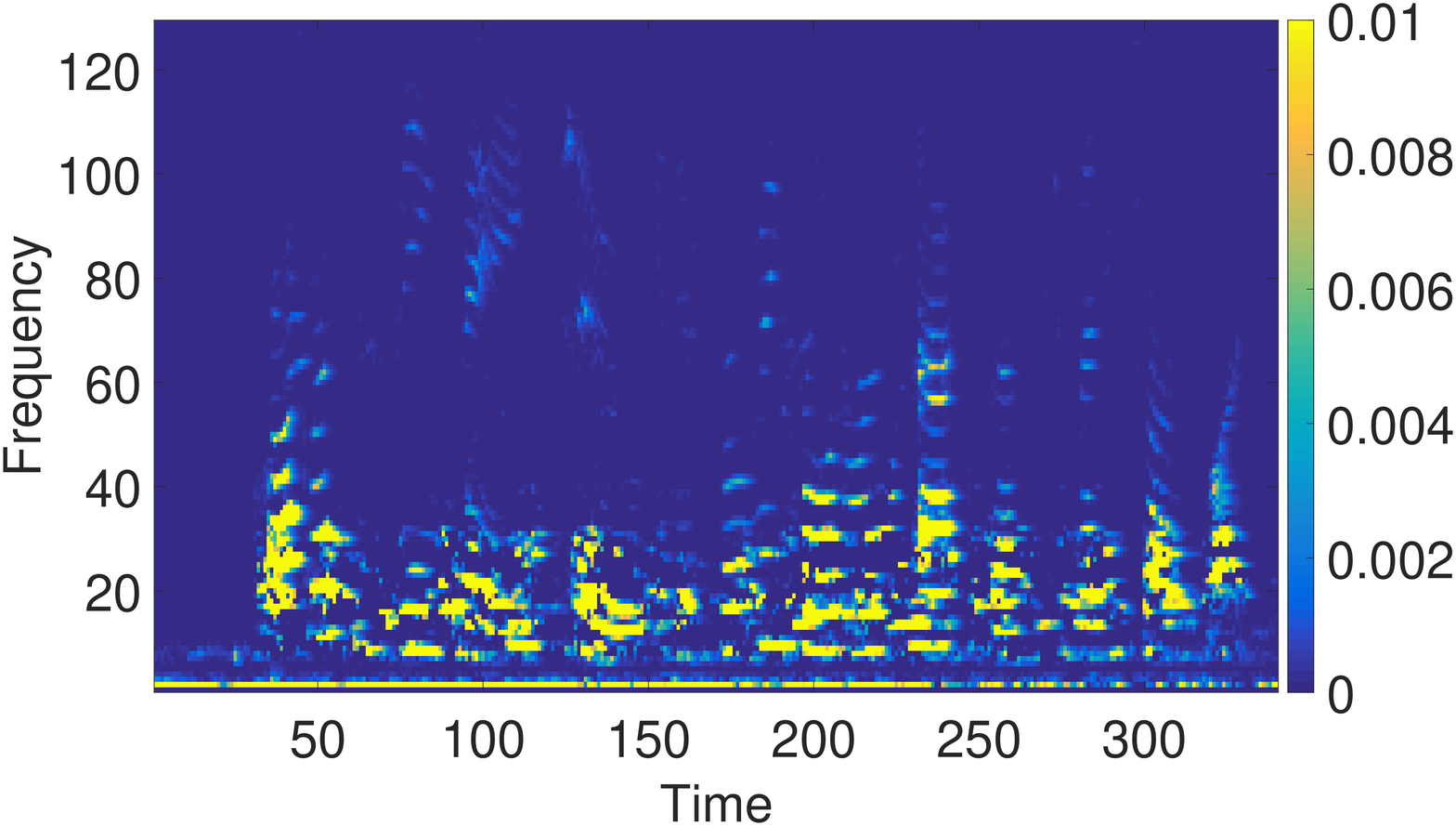}}
  \centerline{(d) Est. PSD (reverb ignored)}\medskip
\end{minipage}
\caption{Estimated PSD for source $1$ in the presence of $4$ concurrent sources in a practical reverberant room.}
\label{fig:lab_room}
\end{figure}
\subsection{In a simulated environment}
\label{sec: simulated-environment}
In the final part of our evaluation, we simulated different reverberant and non-reverberant conditions using image source method \cite{allen1979image,habets2006room}. The clean speech signals were taken from the WSJCAM0 corpus \cite{robinson1995wsjcamo} and a $4^{th}$ order SMA was used in the simulation. The source locations were assumed to be known. Table \ref{table:psd-error} presents the average SIR under different conditions in a room with $[6 \times 7 \times 6]$ m dimension. We ran each simulation $20$ times with random audio signals at random azimuths on the same plane and took the average values. While the performance of the system conceivably depended on the number of the sources for the non-reverberant case, it varied with the reverberation time ($T_{60}$) in a reverberant room. One of the reason for the performance issue in the highly reverberant environment could be due to the violation of the multiplicative transfer function \cite{avargel2007multiplicative} assumption that the impulse response is shorter than the analysis window ($32$ ms). As it is not always practical to increase the analysis window length due to the non-stationarity of the speech signal, a solution could be to model the algorithm using the convolutive transfer function \cite{avargel2007system}. An example of the estimated waveforms is shown in Fig. \ref{fig:simulated_room} for the case of $T_{60} = 0.5$ s. Notably from Fig. \ref{fig:simulated_room}, while the BF partially restored the signal, the estimated PSD-based Wiener post-filter significantly improved the accuracy.\par
\begin{table}[ht]
\caption{Average SIR (dB) based on $20$ simulations in each case}
\label{table:psd-error}
\begin{center}
\renewcommand{\arraystretch}{1.2}
\begin{tabular}{|l|c|c|c|}
\hline
\multirow{2}{*}{\textbf{Non-reverberant}} & $L=4$ & $L=6$ & $L=8$ \\ \cline{2-4}
& $25.67$ & $16.98$ & $10.58$ \\
\hline
\multirow{2}{*}{\textbf{Reverberant ($L=4$)}} & $T_{60}=.2\text{s}$ & $T_{60}=.3\text{s}$ & $T_{60}=.5\text{s}$ \\ \cline{2-4}
 & $11.04$ & $7.35$ & $4.25$ \\
\hline
\end{tabular}
\end{center}
\end{table}
\begin{figure}[!ht]
\begin{minipage}[b]{0.5\linewidth}
  \centering
  \centerline{\includegraphics[width=\linewidth]{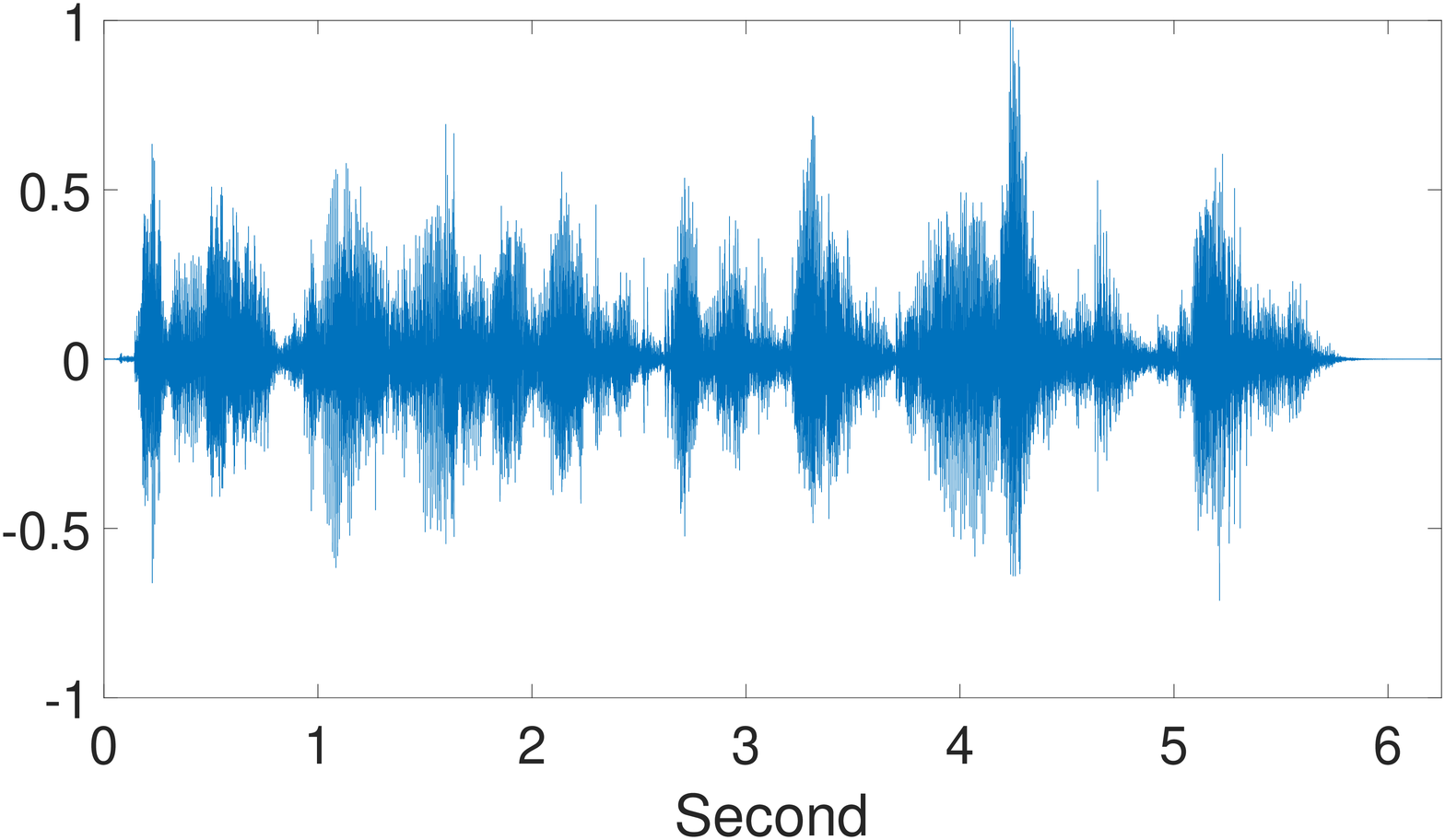}}
  \centerline{(a) Mixed reverberant signal}\medskip
\end{minipage}
\hfill
\begin{minipage}[b]{.5\linewidth}
  \centering
  \centerline{\includegraphics[width=\linewidth]{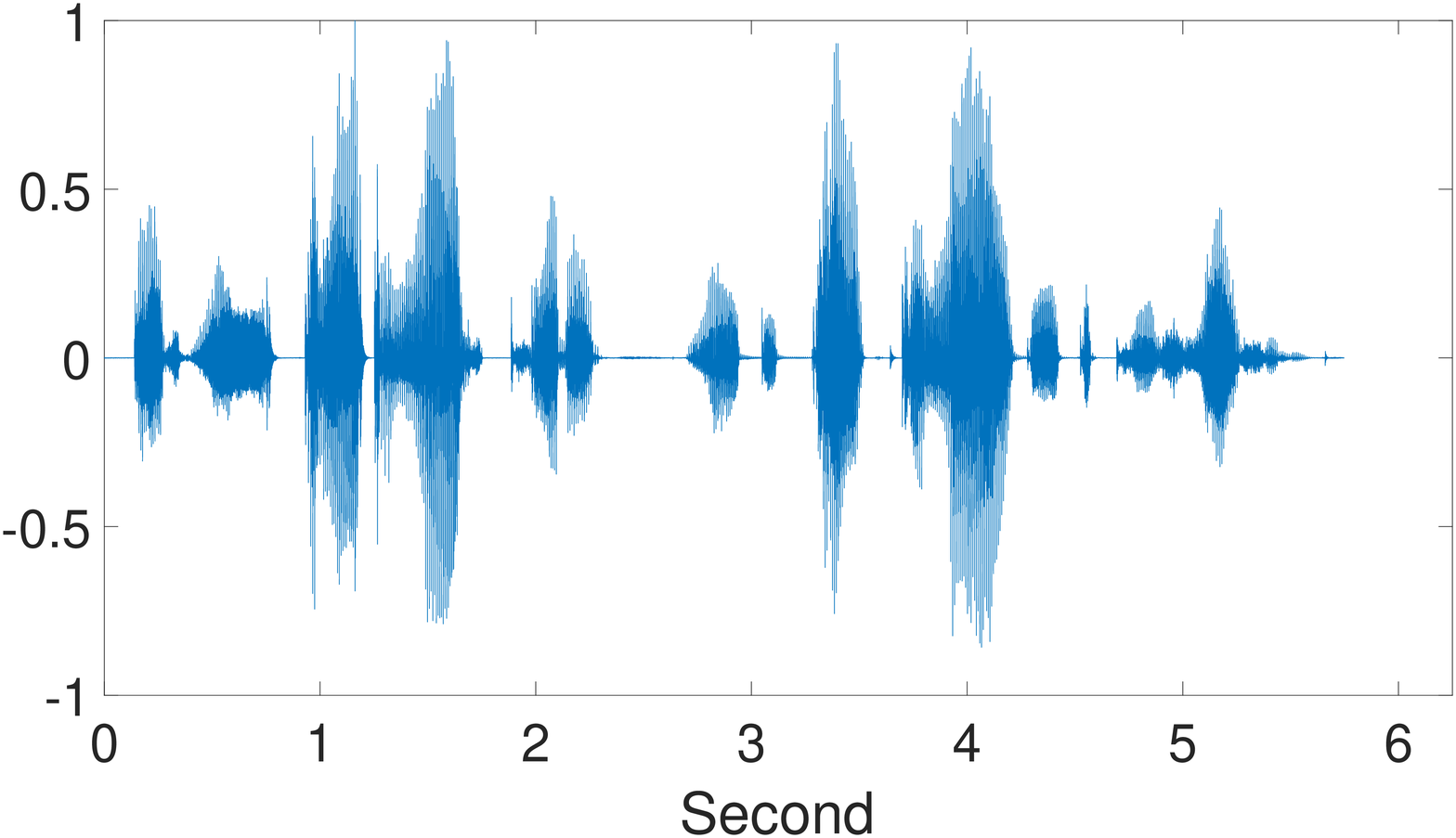}}
  \centerline{(b) Clean signal}\medskip
\end{minipage}
\begin{minipage}[b]{0.5\linewidth}
  \centering
  \centerline{\includegraphics[width=\linewidth]{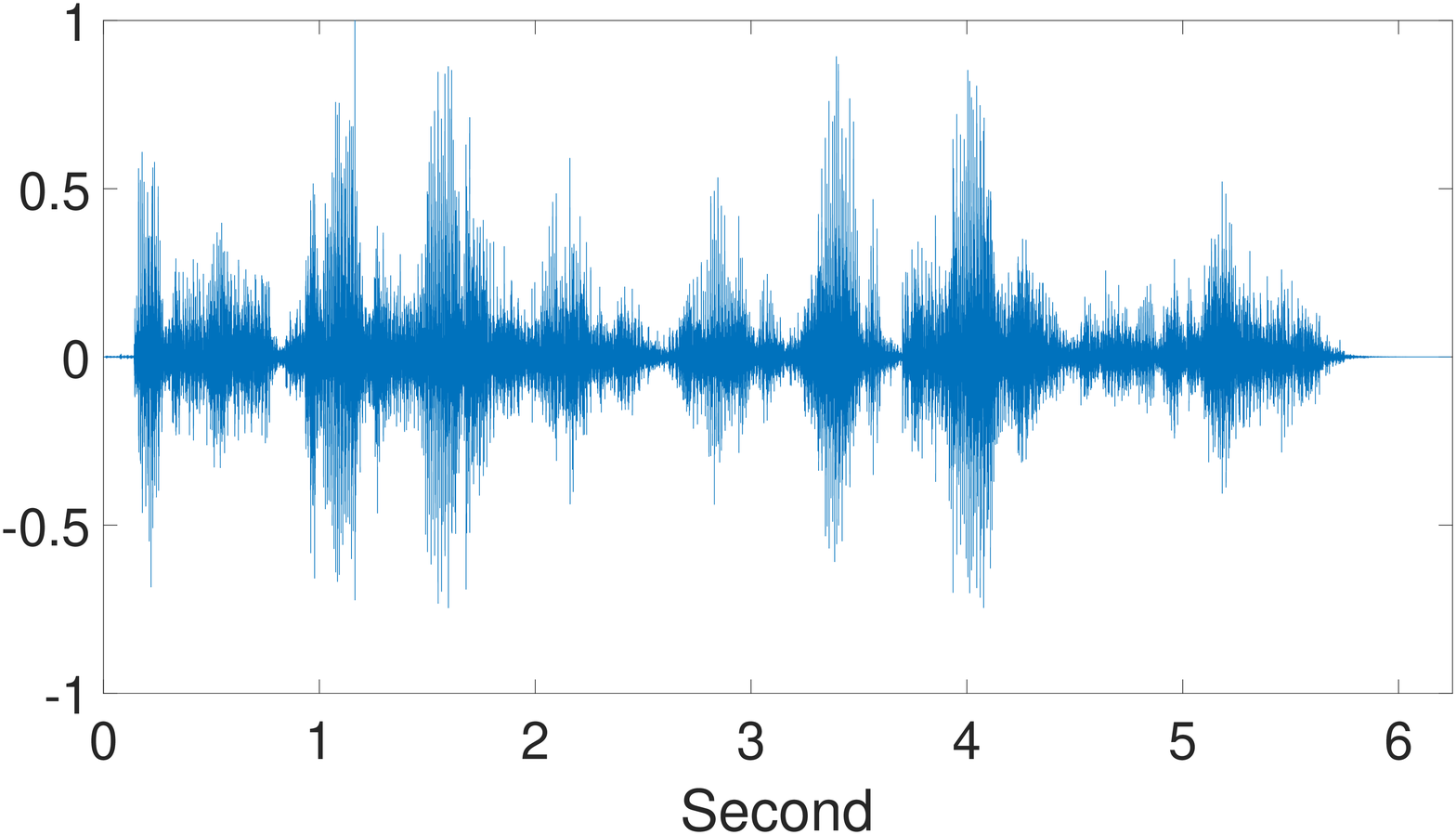}}
  \centerline{(c) Beamformer output}\medskip
\end{minipage}
\hfill
\begin{minipage}[b]{.5\linewidth}
  \centering
  \centerline{\includegraphics[width=\linewidth]{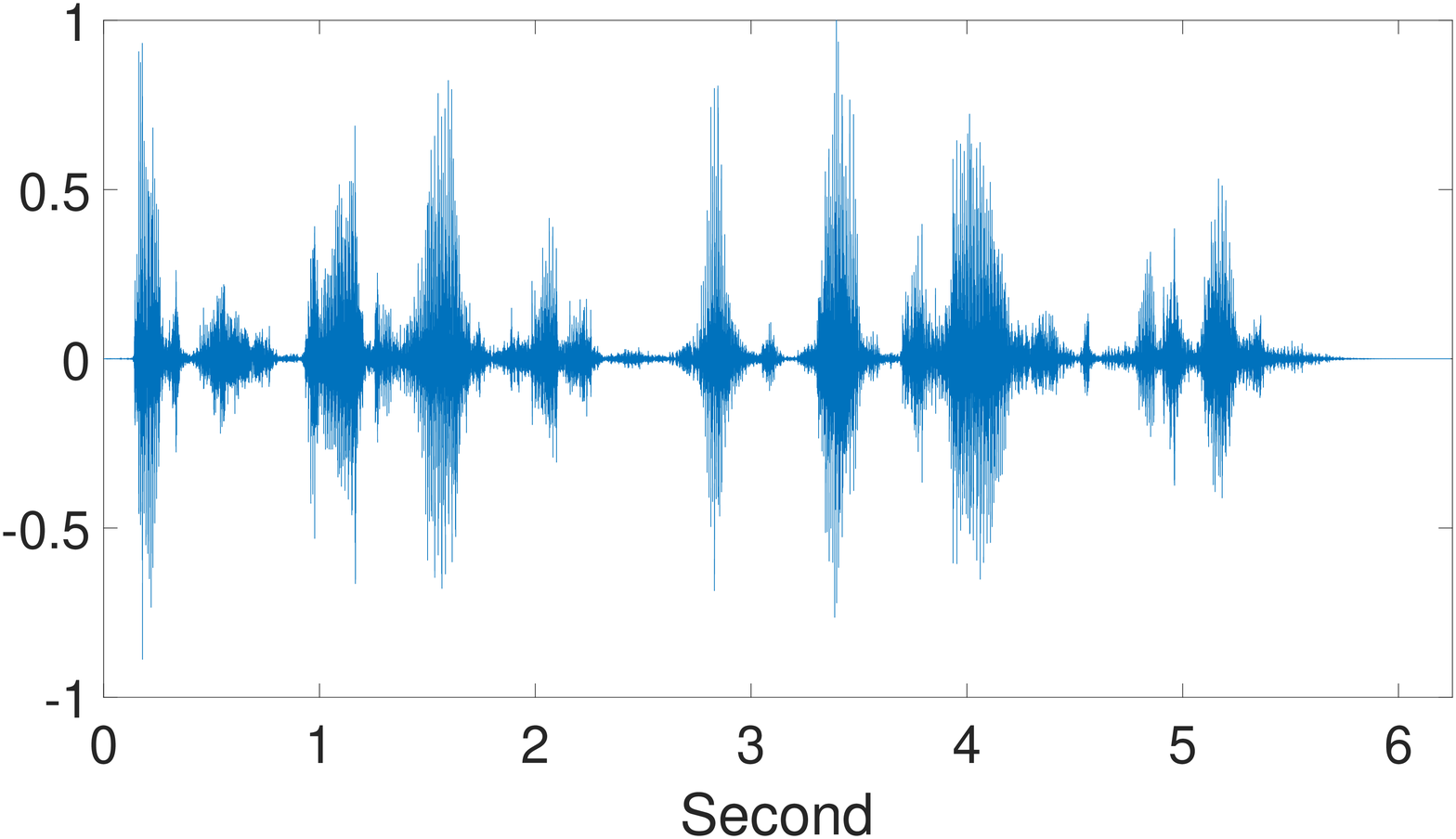}}
  \centerline{(d) Final output}\medskip
\end{minipage}
\caption{An example of the estimated waveform in a simulated reverberant room with $T_{60}=0.5$ s and $L=4$.}
\label{fig:simulated_room}
\end{figure}
\section{Conclusion}
\label{sec:conclusion}
We proposed a method to estimate the PSDs of multiple sources in a reverberant room. The algorithm was formulated in the SH domain to take the advantage of the orthogonality of the SH. We demonstrated an application of the proposed method by performing source separation in distinct multi-source scenarios. The end-to-end performance was evaluated using an Eigenmike under different practical and simulated environments. The algorithm showed satisfactory performance in terms of different objective evaluations for PSD estimation and source separation. For future work, we plan to investigate the performance of the algorithm in a noisy reverberant room with high $T_{60}$.
% -------------------------------------------------------------------------
% Either list references using the bibliography style file IEEEtran.bst
\bibliographystyle{IEEEtran}
\bibliography{refs17}
%
% or list them by yourself
% \begin{thebibliography}{9}
% 
% \bibitem{waspaa17web}
%   \url{http://www.waspaa.com}.
%
% \bibitem{IEEEPDFSpec}
%   {PDF} specification for {IEEE} {X}plore$^{\textregistered}$,
%   \url{http://www.ieee.org/portal/cms_docs/pubs/confstandards/pdfs/IEEE-PDF-SpecV401.pdf}.
%
% \bibitem{PDFOpenSourceTools}
%   Creating high resolution {PDF} files for book production with 
%   open source tools, 
%   \url{http://www.grassbook.org/neteler/highres_pdf.html}.
%
% \bibitem{eWilliams1999}
% E. Williams, \emph{Fourier Acoustics: Sound Radiation and Nearfield Acoustic
%   Holography}. London, UK: Academic Press, 1999.
% 
% \bibitem{ieeecopyright}
%   \url{http://www.ieee.org/web/publications/rights/copyrightmain.html}.
%
% \bibitem{cJones2003}
% C. Jones, A. Smith, and E. Roberts, ``A sample paper in conference
%   proceedings,'' in \emph{Proc. IEEE ICASSP}, vol. II, 2003, pp. 803--806.
% 
% \bibitem{aSmith2000}
% A. Smith, C. Jones, and E. Roberts, ``A sample paper in journals,'' 
%   \emph{IEEE Trans. Signal Process.}, vol. 62, pp. 291--294, Jan. 2000.
% 
% \end{thebibliography}

\end{sloppy}
\end{document}